\documentclass[%
 reprint,
superscriptaddress,
amsmath,amssymb,
pra,
]{revtex4-2}
\usepackage{amsmath,amssymb}
\usepackage{graphicx,float,color}
\usepackage{algorithmicx,algorithm,algpseudocode}
\usepackage{makecell} 
\usepackage{ulem} 
\floatname{algorithm}{Procedure}
\usepackage{subcaption}
\usepackage{hyperref}
\usepackage{xcolor}
\def\bI{\mathbb{I}}
\def\bZ{\mathbb{Z}}

\def\cC{\mathcal{C}}
\def\cE{\mathcal{E}}

\def\cN{\mathcal{N}}
\def\cO{\mathcal{O}}
\def\cP{\mathcal{P}}
\def\cS{\mathcal{S}}
\def\cT{\mathcal{T}}

\def\QEC{\mathsf{QEC}}

\def\ol{\overline}
\def\ot{\otimes}
\def\ths{\thickspace}
\def\ts{\thinspace}
\def\Tr{\mathsf{tr}}

\newcommand{\figsdir}{./}
\newcommand{\diagramdir}{./}
\newcommand{\predictscale}{0.8}
\newcommand{\diagscale}{0.1}
\newcommand{\predvspace}{0.3cm}

\definecolor{blue}{rgb}{0.5, 0, 0.13}

\newcommand\soutpars[1]{\let\helpcmd\sout\parhelp#1\par\relax\relax}
\long\def\parhelp#1\par#2\relax{%
  \helpcmd{#1}\ifx\relax#2\else\par\parhelp#2\relax\fi%
}

\date{\today}
\begin{document}

\title{Improved quantum error correction with randomized compiling}

\author{Aditya Jain}
\affiliation{
Department of Applied Mathematics, University of Waterloo, Waterloo, Ontario N2L 3G1, Canada.
}
\affiliation{%
Institute for quantum computing, University of Waterloo, Waterloo, Ontario N2L 3G1, Canada.
}
\affiliation{
Keysight Technologies Canada Inc., Mississauga, Ontario L5N 2M2, Canada.
}

\author{Pavithran Iyer}
\affiliation{
Department of Applied Mathematics, University of Waterloo, Waterloo, Ontario N2L 3G1, Canada.
}

\affiliation{%
Institute for quantum computing, University of Waterloo, Waterloo, Ontario N2L 3G1, Canada.
}

\affiliation{%
Xanadu, Toronto, Ontario M5G 2C8, Canada. 
}

\author{Stephen D. Bartlett}
\affiliation{%
Centre for Engineered Quantum Systems, School of Physics, University of Sydney, Sydney, New South Wales 2006, Australia.
}

\author{Joseph Emerson}
\affiliation{
Department of Applied Mathematics, University of Waterloo, Waterloo, Ontario N2L 3G1, Canada.
}
\affiliation{%
Institute for quantum computing, University of Waterloo, Waterloo, Ontario N2L 3G1, Canada.
}
\affiliation{
Keysight Technologies Canada Inc., Mississauga, Ontario L5N 2M2, Canada.
}

\begin{abstract}
Current hardware for quantum computing suffers from high levels of noise, and so to achieve practical fault-tolerant quantum computing will require powerful and efficient methods to correct for errors in quantum circuits. Here, we explore the role and effectiveness of using noise tailoring techniques to improve the performance of error correcting codes. Noise tailoring methods such as randomized compiling (RC) convert complex coherent noise processes to effective stochastic noise. While it is known that this can be leveraged to design efficient diagnostic tools, we explore its impact on the performance of error correcting codes. Of particular interest is the important class of coherent errors, arising from control errors, where RC has the maximum effect -- converting these into purely stochastic errors. For these errors, we show here that RC delivers an improvement in performance of the concatenated Steane code by several orders of magnitude. We also show that below a threshold rotation angle, the gains in logical fidelity can be arbitrarily magnified by increasing the size of the codes. These results suggest that using randomized compiling can lead to a significant reduction in the resource overhead required to achieve fault tolerance.
\end{abstract}

\maketitle

\section{Introduction}
\label{sec:introduction}
Noise is pervasive in present-day quantum computation. The theory of fault tolerance was developed to guarantee reliable computations in the presence of noise. However, fault tolerant constructions demand a large overhead in terms of additional resources required to encode a logical computation in a way that is resilient to errors. Achieving the target logical error rates as required by various applications with the limited amount of resources in terms of the number of physical qubits is a challenging task. Along with designing better error correcting codes, decoders and high quality hardware components of a quantum computer, there are other ways of reducing logical error rates. Active noise tailoring by randomized compiling (RC) \cite{WE16} is a potential candidate for two key reasons. First, RC significantly simplifies the form of the noise on the encoded quantum information. Second, RC can be used to transform an unknown error model into one that is adapted to the error correction capabilities of a particular code.

Randomized compiling tools were leveraged to accurately predict the performance of quantum error correction schemes in Ref.~\cite{Iyer2021}. Although simplifying the form of the noise makes the performance more predictable, it was observed that RC can sometimes degrade the performance of an error correcting code. We can understand this effect by using the $\chi$-representation \cite{WBC15} of a physical noise process. In this representation, the action of noise on a quantum state $\rho$ is given by: $\cE(\rho) = \sum_{i,j}\chi_{i,j}P_{i}\rho P_{j}$ where $P_i$ denote Pauli matrices in the $n-$qubit Pauli group $\cP_{n}$ without phases, i.e., $P_i \in \cP{n}/\{\pm 1, \pm i\}$. Noise tailoring methods such as RC can transform the elements of the $\chi$-matrix, for example by removing off-diagonal elements $\chi_{i,j} \ths \forall \ths i \neq j$. This mathematical transformation is commonly referred to as twirling \cite{BDP96,BBP96,CB2019}.
If one were to remove the contribution of $\chi_{i,j}$ corresponding to Pauli errors that are correctable by the decoder, this could have a negative impact of the code's performance. In general, noise tailoring methods are oblivious to the details of what error terms are relevant for quantum error correction.

The impact of twirling the noise on the performance of error correction schemes has been explored in the literature under various settings. The performance of surface codes under coherent and incoherent error models have been compared in Ref.~\cite{BEK18}, and using numerical studies it was noted that while the threshold is similar in both cases, the subthreshold performance of the twirled channel is significantly better than the original coherent error model. In another setting, analytical calculations of the logical error rate of repetition codes under rotation errors reveal that coherent errors can accumulate faster, leading to worse logical error rates than their corresponding Pauli approximations~\cite{GD17}. The necessity of active coherence-suppression methods for codes with large distances was also noted, but their impact on the code’s performance was not explored. For the Toric code under coherent error models, a laborious analysis has shown that the effective logical channel approaches an incoherent channel provided the noise decreases with increasing code size~\cite{Iverson2020}. However, in the scenario where the error rate remains constant independent of the code size, there are several challenges to arriving at a similar conclusion. In Ref.~\cite{GSCL16}, the poor predictability of the logical error rate and the code’s pseudo threshold under coherent errors provided by their twirled counterparts was identified, reinforcing the need for active noise tailoring. The impact of twirling the noise for complex error models, such as combinations of stochastic errors and rotations around an arbitrary non-Pauli axis, is unknown. The scaling of the potential gains from twirling with increased code-concatenation levels remains unexplored.

In this paper, we analyze the impact of RC on the performance of quantum error correction. In particular, we show that RC improves the performance of a concatenated Steane code under a coherent noise model (specifically, a tensor product of arbitrary identical unitary errors). This positive result demonstrates that RC tools can play a key role in achieving fault tolerance. We present a detailed study of the performance gains with respect to changes in the axis of rotation and the number of levels of concatenation. We identify a special axis of rotation for a given concatenation level where maximum gains from RC are achieved. We note that this axis can be different from the axes of rotation for which the best pseudo-threshold for the code is achieved. It has been observed, in previous studies, that randomized compiling can also degrade logical performance~\cite{BWGB18}. Our study shows that a wide class of physically motivated error models do not exhibit such behaviour. However, we identify some complex noise models where such degradation can occur and provide numerical results for the same.

The paper is structured as follows. In section \ref{sec:background}, we introduce the necessary background material including noise processes, randomized compiling and quantum error correction. Section \ref{sec:methods} discusses the methods used to study the impact of randomized compiling on the logical performance. In section \ref{sec:results}, we present analytical studies for gains offered by randomized compiling using realistic error models. Finally, in section \ref{sec:conclusion} we provide concluding remarks and describe some interesting open problems. 

\section{Background}
\label{sec:background}
In this section, we review the mathematical description of noise processes in quantum circuits as well as the formalism of stabilizer quantum error correction. 

\subsection{Noise in quantum circuits}

The interaction of a quantum system with its environment manifests as errors on the stored quantum information. While the system and its environment together undergo unitary time evolution, the system's reduced dynamics is often a non-unitary map. Markovian noise processes are described by completely positive trace preserving (CPTP) maps $\cE: \rho \mapsto \cE(\rho)$. One of the common ways to represent a CPTP map is using the $\chi$-matrix: $\chi(\cE)$, a $4\times 4$ matrix where: $\cE(\rho) = \sum_{i,j}\chi_{i,j} P_{i} \rho P_{j}$, where $P_{i}, P_{j}$ are Pauli matrices.

A special subclass of noise processes that are widely analyzed in developing fault-tolerant protocols is Pauli channels. They correspond to the probabilistic action of Pauli matrices on the input state, i.e., $\cE(\rho) = \sum_{i,j}\chi_{i,i} P_{i} \rho P_{i}$, where $\chi_{i,i}$ can be interpreted as the probability of the Pauli error $P_{i}$.

While it is easy to study quantum error correction on Pauli error models, unfortunately realistic noise is often poorly approximated by Pauli error models. This causes a severe disparity between error models that can be accurately analyzed in theory and those that occur in experiments. Noise tailoring, achieved through Randomized compiling \cite{WE16}, is a promising tool that helps resolve this disparity. With RC, the average logical performance of a QEC scheme over several compilations with random Pauli gates can be well approximated by the performance of the QEC scheme under an effective Pauli error model. The effective Pauli error model is nothing but the Pauli twirl of the underlying CPTP noise process $\cE$, denoted by $\cT(\cE)$ defined as
\begin{gather}
\cT(\cE)(\rho) = \sum_{P \in \cP_{n}} P \cE(P \rho P) P \label{eq:twirl_channel} \ths .
\end{gather}
We will use the notation $\cE^{T}$ to denote the Pauli Twirl of the CPTP map $\cE$: $\cT(\cE)$.

\subsection{Quantum Error Correction}

An $[[n,k]]$ stabilizer code $\cC$ is a $2^{k}$ dimensional space defined as: $\cC = \{ |\psi\rangle \ts : \ts S_{i}|\psi\rangle = |\psi\rangle \ts , \ts 1 \leq i\leq n-k\}$, where $S_{i}$ are stabilizer generators. See Ref.~\cite{GotPhd97} for an introduction to stabilizer codes and fault tolerance. Concatenated codes are a family of codes where we encode the physical qubits at level $\ell$ using the code at level $\ell-1$. This is a way of constructing larger codes from smaller ones and these codes are typically used to guarantee error suppression in fault tolerance proofs \cite{AGP07,CTV17}.

Measuring stabilizer generators yields a signature of the error that occurred called a syndrome. Inferring the error from the syndrome is called decoding. There are several ways to define a decoder, the simplest of which is the minimum weight decoder. It selects a Pauli error of minimum Hamming weight consistent with the observed syndrome. While some errors on the encoded states can be undone by quantum error correction, there are uncorrectable errors that cause unwanted logical operations on the encoded states under a quantum error correction routine. These uncorrectable errors determine the logical error rate. A valuable tool to define the logical error rate is the effective channel, which encapsulates the effect of a physical noise process and a quantum error correction protocol on the encoded quantum information.

Besides the error-correcting code and the underlying physical noise process, the effective channel is a function of the measured error syndrome $s$. We will use the notation $\cE^{s}_{1}$ to denote the effective channel where the subscript ``1" refers to one encoding level. The relevance of the subscript becomes crucial for concatenated codes \cite{GotPhd97}, where $\cE^{s}_{\ell}$ refers to the effective channel for a level$-\ell$ concatenated code. A particularly useful quantity is the average of logical channels $\cE^{s}_{\ell}$ over all syndrome outcomes, denoted by $\ol{\cE}_{\ell}$:
\begin{gather}
\ol{\cE}_{\ell} = \sum_{s}\cE^{s}_{\ell}\Pr(s) \ths , \label{eq:avgchan}
\end{gather}
where $\Pr(s)$ is the probability of observing the outcome $s$ \cite{IP17,BEK18,CWBL17}. The average logical channel $\ol{\cE}_{\ell}$ indicates how quantum error correction suppresses the effect of physical errors, on average. We will use logical infidelity $r(\ol{\cE}_{\ell})$ \cite{IP17,GSCL16} as a measure of the logical error rate.

The average logical infidelity for a code under a noise process $\cE$ is calculated using the following equation: \cite{Iyer2021}
\begin{gather}
r(\ol{\cE}_{1}) = 1 -  \sum_{\substack{E,E^{\prime} \in \cE_{C} \\ s(E) = s(E^{\prime}) \ts , \ts \ol{E} = \ol{E}^{\prime}}} \phi(E) \ths  \phi^{\star}(E^\prime) \ths \chi_{E,E^{\prime}}  \ths ,\label{eq:loginfidexp}
\end{gather}
where $\chi_{i,j}$ represents the $(i,j)^{th}$ entry of the $\chi-$matrix of $\cE$, $\cE_{C}$ is the set of correctable errors, $\ol{E}$ is the logical component in the decomposition of $E$ with respect to the Stabilizer group and $\phi(E)$ is specified by $R_{s(E)} E = \phi(E) \ths S$ for any Pauli error $E$ and some stabilizer $S$. We use this expression at various points to calculate the logical infidelity. To calculate the entries of the $\chi-$matrix of the effective logical channel we use the following general expression: \cite{Iyer2021}
\begin{gather}
\chi(\ol{\cE}_{1})_{l,m} = \sum_{\substack{E,E^{\prime} \in \cE_{C} \\ s(E) = s(E^{\prime}) \ts , \ts \ol{E} = \ol{E}^{\prime}}} \!\!\! \phi(E, l) \ths  \phi^{\star}(E^\prime, m) \ths \chi_{E \ol{P}_{l},\ol{P}_{m}E^{\prime}} \ths .\label{eq:chi_lm_log_phys}
\end{gather}
where $R_{s(E)} \ths |E \ths \ol{P}_{l}| = \phi(E,l) \ths S \ths |\ol{P}_{l}|$, for $l \in \{0,1,2,3\}$, any Pauli error $E$ and some stabilizer $S$. Here $|P|$ stands for the bare Pauli without any associated global phase. 

We calculate the $\chi-$matrix for logical channels at higher levels i.e, for $\ell>1$ by recursing the expression in Eq.(\ref{eq:chi_lm_log_phys}) and using the entries of $\chi(\ol{\cE}_{\ell-1})$ in the right hand side to evaluate $\chi(\ol{\cE}_{\ell})$.

\section{Methods}
\label{sec:methods}

The goal of this paper is twofold. First, we want to identify important scenarios for physical errors wherein RC can be leveraged to improve the performance of quantum error correcting codes. Second, identify settings under which such performance gains cannot be guaranteed. For the first goal, we study the performance of concatenated Steane code under realistic error models. We start off by simple rotations about $Z-$axis and progressively move to arbitrary rotations followed by a combination of coherent and stochastic error models. For the second goal, we generate numerical results for a large ensemble of noise processes belonging to more complex noise models which involve random rotations on different qubits and arbitrary CPTP maps. All the performance metrics in this paper are derived in the memory model and assume perfect syndrome extraction. Simulations with gate dependent errors can be pursued in the future.

For both the goals, it is crucial to understand how RC can be applied alongside quantum error correction in practice. We follow the methods of Ref.~\cite{Iyer2021}. The main idea can be summarized as follows. Recall that noise tailoring by randomized compiling is achieved by inserting random Pauli gates in a circuit such that its net effect does not change the logical output of the circuit. Consequently, the average output distribution of the circuit over all possible Pauli random gates can be understood by studying the response of the original circuit against Pauli noise on the individual components. In the same spirit, we insert random Pauli gates around all the individual components of a quantum error correction circuit. There is no need to account for sources of noise in the extra Pauli random gates because they can be absorbed into the original elements of the quantum error correcting circuit. The theory of RC prescribes exponentially many compilations of the underlying circuit to achieve perfect twirling. However, in practice, only a handful compilations can be realized~\cite{Hashim2021}. Despite this practical limitation, we assume the ideal application of RC in this paper for simplicity. We leave the details of this procedure to the appendix section \ref{sec:app_qec_rc}. 

We now have two variations of the average fidelity. First, the standard notion -- average fidelity over all syndrome outcomes, $r(\ol{\cE}_{1})$, defined in Eq.~\eqref{eq:loginfidexp}. Second, the average fidelity over syndrome outcomes as well as logically equivalent compilations of the quantum error correction circuit, which we will denote $r_{\mathrm{rc}}$. Note that the number of random compilations for a circuit with $n$ elements grows as $\cO(4^{n})$. In the ideal case, where we have considered all of these compilations in $r_{\mathrm{rc}}(\ol{\cE}_{1})$, it reduces to $r(\ol{\cE}^{T}_{1})$.

While Eq.~\eqref{eq:loginfidexp} addresses the logical channel of a block code, we can easily extend these definitions for a concatenated code assuming a hard decoder \cite{CWBL17,GSCL16}. In this case, the logical channel at level$-\ell$ can be recursively defined in as a function whose input physical channels are the logical channels at level$-(\ell-1)$. We will use the notation $r(\ol{\cE}_{\ell})$ and $r(\ol{\cE}^{T}_{\ell})$ to denote the logical channels of a level$-\ell$ concatenated code without RC and with RC, respectively. Their ratio, denoted by $\delta_{\ell}$, where
\begin{equation}
\delta_\ell = \frac{r(\ol{\cE}_{\ell})}{r(\ol{\cE^T}_{\ell})} \ths, \label{eq:deltaDef}
\end{equation}
is an indicator of the performance gain due to RC, which we will estimate for various error models. Note that $\delta_{\ell} > 1$ indicates a performance gain whereas $\delta_{\ell} < 1$ denotes a performance loss.

\section{Results and discussion}
\label{sec:results}

This section is devoted to case studies of performance gains from RC for the concatenated Steane code, under various interesting classes of error models, and inferences we can draw from these studies. Markovian errors can be broadly classified into unital and non-unital maps. Since non-unital components of a noise map do not impact the error rate significantly \cite{JW2015,GD17}, we restrict our attention to unital maps in this paper. In particular, we choose coherent rotations which form an important class of unital maps. In practice, these typically arise from imperfect pulses used to implement quantum gates in the hardware. Interestingly, these are also the class of errors on which randomized compiling has the maximum effect of turning them into purely incoherent noise. 

\subsection{Rotation about  \texorpdfstring{$Z-$}{Lg}axis}
While we ideally want to study the impact of RC on the performance of a quantum error correcting code under general coherent errors, let us first start with a simple yet interesting model -- rotations about the $Z-$axis. Although the RC process tailors the underlying physical noise, irrespective of the choice of the code, through this example we show that in fact the gains produced from RC can be arbitrarily increased by choosing codes of increasing distances.

Recall that the rotation about $Z-$axis is specified by $\rho \rightarrow R_{Z}(\omega)\rho R_{Z}(-\omega)$ where
\begin{gather}
R_{Z}(\omega) = \cos (\omega/2) \ths I + i \sin (\omega/2) \ths Z \ths . \label{eq:rtz}
\end{gather}
Applying the rotation independently across all $n = 7$ the physical qubits of the Steane code, is specified by the map
\begin{equation}
\cE(\bar{\rho}) = R^{\ot n}_{Z}(\omega) \ths \bar{\rho} \ths R^{\ot n}_{Z}(-\omega)
    \label{eq:enc_rtz}.
\end{equation}
The performance of the Steane code under the above error model, can be inferred from Eq.~\eqref{eq:loginfidexp}, where the correctable errors $\cE_{\cC}$ can be defined with respect to the minimum weight decoder. Explicitly enumerating all correctable errors, we find that there are $22$ correctable errors of weight at most one, and $42$ two-qubit ones. Since we are confined to rotations about the $Z-$axes, we can limit ourselves to the correctable errors of $Z-$type. Reserving the details of our derivation to Appendix~\ref{sec:app_log_fid}, we find
\begin{gather}
r(\ol{\cE}_{1}) \approx 63 \ths (\omega/2)^4 - 476 \ths (\omega/2)^6 + \mathcal{O}(\omega^8) \ths . \label{eq:loginfid_Z_rot_noRC}
\end{gather}
In comparison, the logical infidelity for quantum error correction with randomized compiling is
\begin{gather}
r(\ol{\cE^T}_{1}) \approx  21 \ths (\omega/2)^4 - 112 \ths (\omega/2)^6 + \mathcal{O}(\omega^8) \ths . \label{eq:loginfid_RC}
\end{gather}
Finally, the performance gain from RC quantified using the metric $\delta_{1}$ defined in eq. \ref{eq:deltaDef} can now be estimated as
\begin{gather}
\delta_1 = \frac{r(\ol{\cE}_{1})}{r(\ol{\cE^T}_{1})} \approx 3 - \frac{5}{3} \ths (\omega)^2 + \mathcal{O}(\omega^4) \ths . \label{eq:gainL1}
\end{gather}

\begin{figure}[ht]
\centering
\includegraphics[scale=0.6]{\figsdir/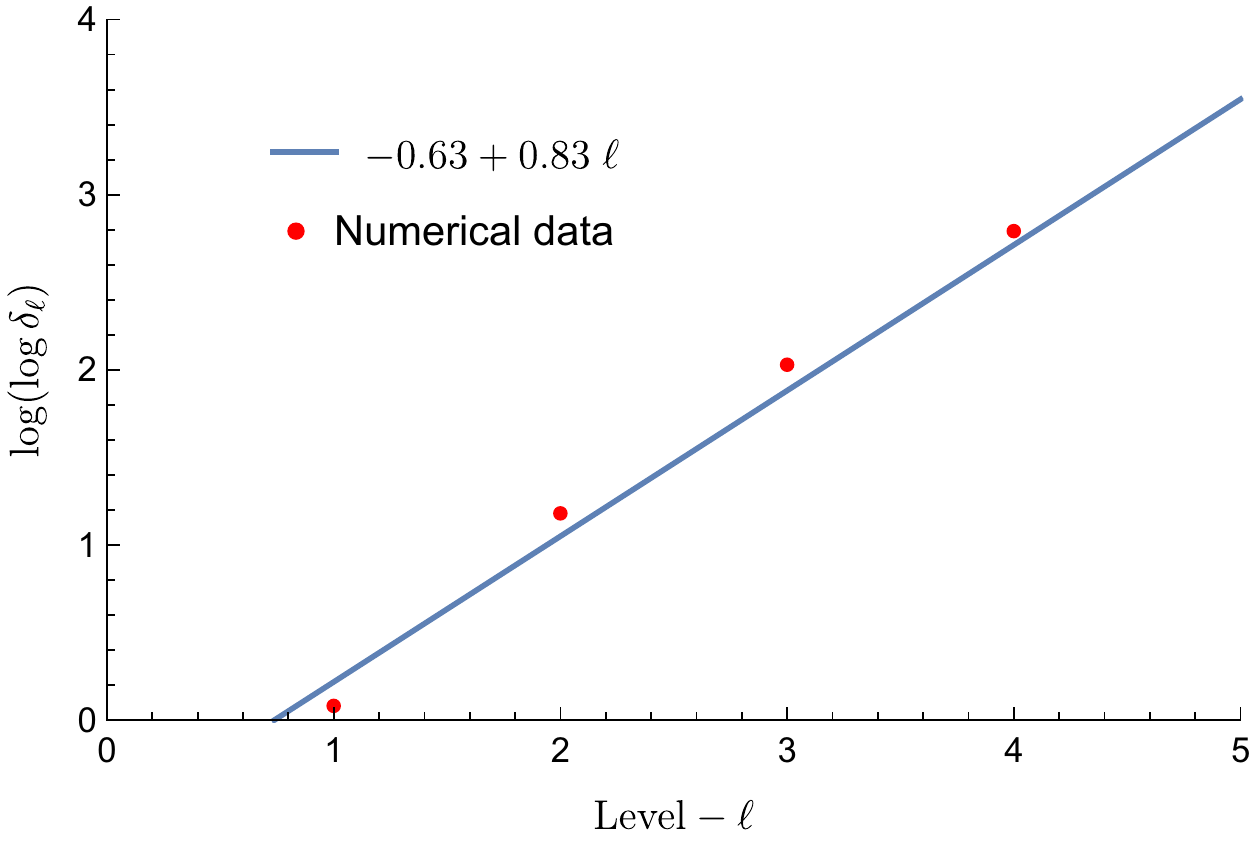}
\caption{The above figure shows that the gain at level $\ell$ and the gain $\delta_{\ell}$ scales doubly exponentially with $\ell$. The rotation angle used here is $\omega=\pi/20$. }
\label{fig:asymprtZ}
\end{figure}

\begin{figure}[ht]
\centering
\includegraphics[scale=0.55]{\figsdir/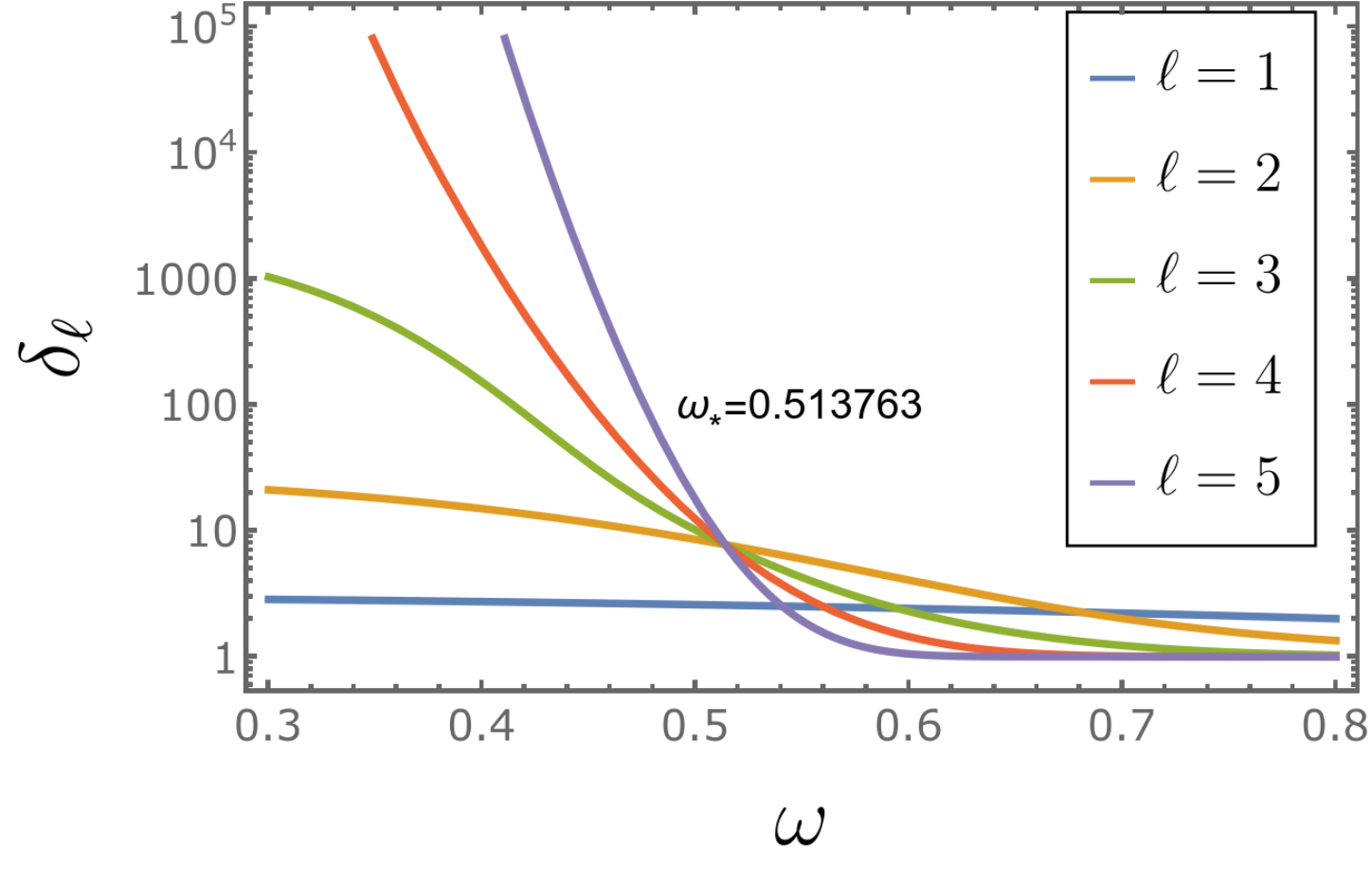}
\caption{Gains in logical performance, $\delta_{\ell}$, of a level $\ell$ concatenated Steane code for rotations by angle $\omega$ about the $Z-$axis. The common crossover point lies at $\omega_\star \approx 0.51$, which corresponds to a rotation angle of about $15^{\circ}$, below which gains from RC can be amplified by increasing the number of levels of concatenation.}
\label{fig:rtZthres}
\end{figure}

We now show that the above modest performance gains can be made arbitrarily large by  concatenating the Steane code with itself. It is possible to extend the analysis above via recursion to approximate the effective logical channel for a level $\ell$ concatenated Steane code for $\ell > 1$. The details of this procedure can be found in Appendix~\ref{sec:app_chilogapp}. The approximate logical channel allows us to estimate the performance of level $\ell$ concatenated Steane code and study the impact of randomized compiling on it.
To understand the impact of RC with the number of levels, we can do a leading order analysis of the recursive relations used to construct the average logical channel, described in Appendix~\ref{sec:app_chilogapp}. We find that for small rotation angle $\omega$, the average infidelity of the logical channel scales as
\begin{align}
r(\ol{\cE}_{\ell}) &\approx 63^{2^\ell - 1} (\omega/2)^{2^{\ell + 1}} \ths , \nonumber\\
r(\ol{\cE^T}_{\ell}) &\approx 21^{2^\ell - 1} (\omega/2)^{2^{\ell + 1}} \ths .
\end{align}
Subsequently, the scaling of gain $\delta_\ell$ with the levels of concatenation is given by
\begin{gather}
\delta_\ell \approx 3^{2^\ell - 1} - (5\times2^{l-1}\times3^{2^l-3}) \omega^2 + O(\omega^4) \ths .
\end{gather}
Figure \ref{fig:asymprtZ} corroborates this scaling law for the exact value of the logical error rates of the concatenated Steane code, in other words, showing that $\log(\log(\delta_\ell)))$ is approximately a linear function of $\ell$. Note that the above analysis is accurate for small rotation angles.
Varying the rotation angles leads us to another important discovery. Figure \ref{fig:rtZthres} shows the gains from randomized compiling for a range of rotation angles for levels $1 \le \ell \le 5$. The gains from RC grow significantly with increase in number of levels of the code. The figure suggests the presence of a threshold rotation angle $\omega_\star$ below which arbitrary gains from RC can be achieved by increasing the size of the code (levels of concatenation). On the contrary, for rotations $\omega > \omega_{\star}$, the trend reverses.

We now turn to more general noise models, where we will find that the presence of a threshold in the case of rotations about the $Z-$axis, extends to the general case.

\subsection{Rotation about an arbitrary axis}

While the above analysis considered coherent error models described by rotations about the $Z-$axis, it is straightforward to apply these ideas to rotations about any of the Pauli axes. We now investigate average gains due to RC for a rotation about an arbitrary axis.

We consider a general error model where the physical qubits of a code undergo rotations about an arbitrary axes of the Bloch sphere, described by the unitary matrix $U$, i.e., $\cE(\bar{\rho}) = U^{\otimes n} \bar{\rho}  (U^{\dagger})^{\otimes n}$. The following parameterization for $U$ \cite{Bartlett2009} is useful for our analysis:
\begin{equation}
\begin{pmatrix}
\cos(\omega/2) + i \sin(\omega/2) \cos(\theta) & i e^{-i\phi} \sin(\omega/2) \sin(\theta)\nonumber\\
i e^{-i\phi} \sin(\omega/2) \sin(\theta) & \cos(\omega/2) - i \sin(\omega/2)
\end{pmatrix}.
\label{eq:arbitUnitary}
\end{equation}
where $0 \le \theta \le \pi$ and $0 \le \phi \le 2 \pi$ define the axis (in polar angles) about which each qubit is rotated, and $\omega$ gives the magnitude of the rotation. For example, $\theta = \phi = 0$ can be identified with rotations about the $Z-$axis. The performance gain from RC can be defined following Eq.~\ref{eq:gainL1}, as a function of the parameters $\delta(\theta, \phi, \omega)$. The average gain for an unknown axis is computed as
\begin{gather}
\ol{\delta}_{\ell}(\omega) = \frac{1}{2 \pi} \int_{0}^{2\pi} d\phi \int_{0}^{\pi} \sin(\theta) \ths d\theta \ths \delta_{\ell}(\theta,\phi, \omega) \ths  , \label{eq:ave_gain_arbitrary_rot}
\end{gather}
for $\ell = 1$. Likewise, for concatenated codes, $\ol{\delta}_{\ell}$ denotes the average gain in performance for level $\ell$. This is similar to the conclusion drawn for the case of rotations about the $Z-$axis. First of all we see that for all coherent errors RC improves the performance of the Steane code. Furthermore, performance gains are largest for coherent errors that correspond to rotations about the $X,Y$ or $Z$ axes.

\begin{figure}[ht]
\centering
\includegraphics[scale=0.8]{\figsdir/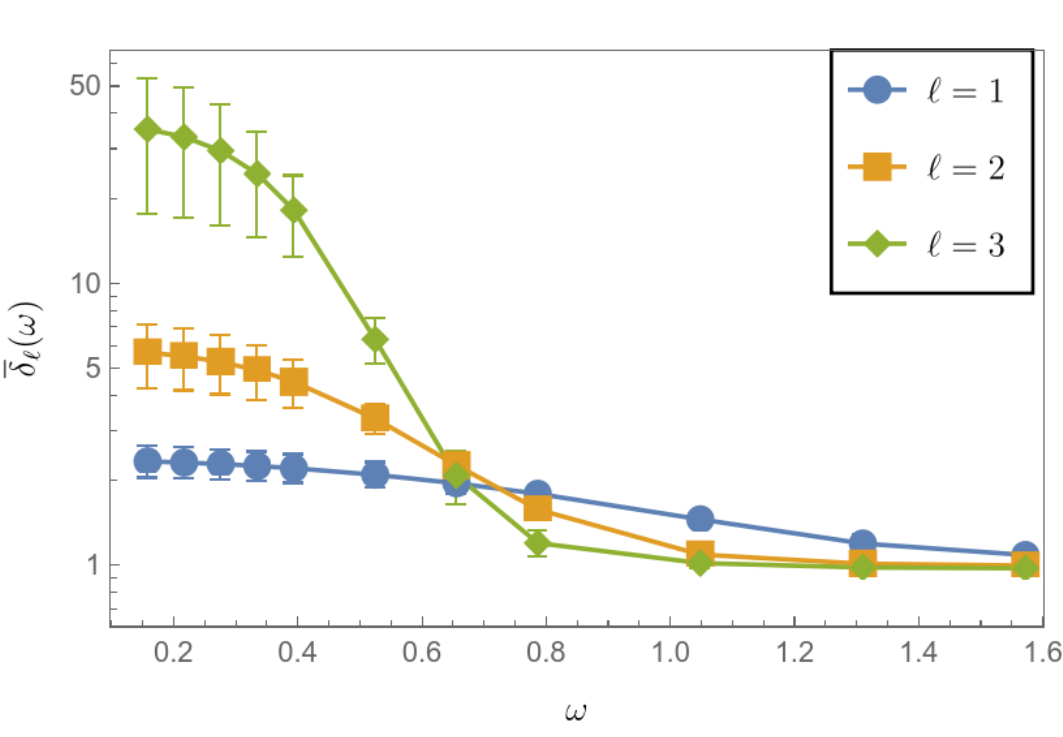}
\caption{The average gain in performance from RC, using the Haar average over all axes of rotation, for the level $\ell$ concatenated Steane code. The average gains are larger for small magnitudes of rotation. We observe that the gains increase significantly with the number of levels for $\omega \le \ol{\omega}_{\star} \approx 0.65$, which corresponds to a rotation angle of about $19^{\circ}$.}
\label{fig:arbitRotHaarThresh}
\end{figure}

\begin{figure}[ht]
\centering
\includegraphics[scale=0.65]{\figsdir/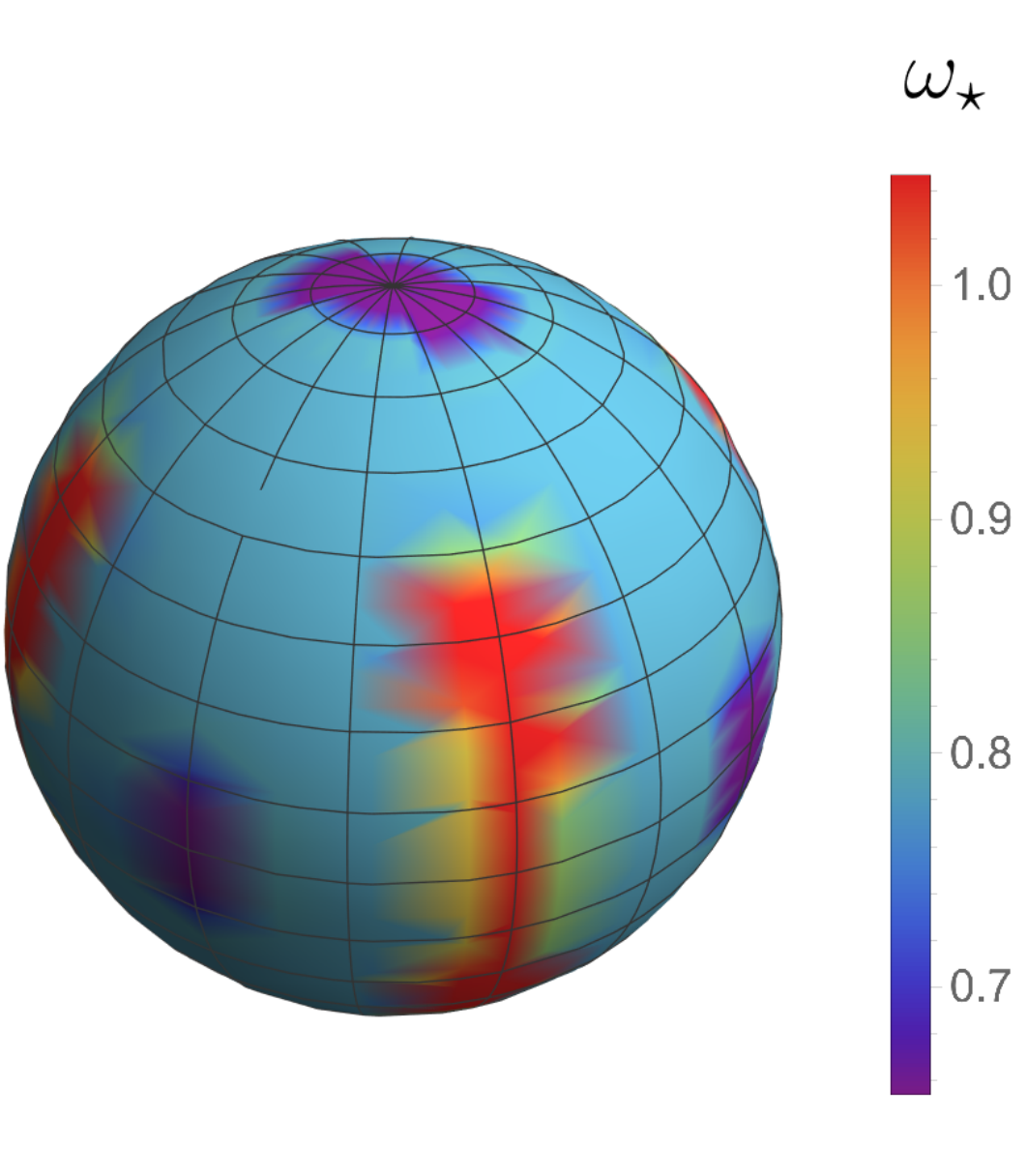}
\caption{The threshold angle $\omega_{\star}$ for which $\delta_2 < \delta_1$ for rotations about an axis parameterized by $\theta, \phi$. Each point on the sphere has coordinates $\{\sin(\theta)\cos(\phi),\sin(\theta)\sin(\phi),\cos(\theta)\}$ and the color denotes the threshold value of angle $\omega$ for which the above condition holds. It shows that the cardinal axes do not have the highest threshold.}
\label{fig:arbitRotHaarPseduoThresh}
\end{figure}

Using the general techniques developed in the appendix to approximate the effective logical channel of a level$-\ell$ concatenated code, we can estimate the gains $\ol{\delta}_{\ell}$ in average performance due to RC over the various rotation axes. Similar to the case of $Z-$rotations, Fig.~\ref{fig:arbitRotHaarThresh} suggests the presence of a threshold $\ol{\omega}_{\star}$ wherein for rotation angles $\omega \leq \ol{\omega}_{\star}$ the gains can be arbitrarily increased by choosing codes of larger distance, whereas the trend reverses for $\omega > \ol{\omega}_{\star}$.
Note that threshold angle $\ol{\omega}_{\star}$ for rotations about an unknown axis is higher the threshold for rotations about the $Z-$axes, i.e., $\ol{\omega}_{\star} > \omega_{\star}$. This can be explained as follows. In the case of a generic non-Pauli axis, the twirled noise model, i.e., is in the presence of RC, is composed of a probabilistic mixture of $X, Y$ and $Z$ type errors. Whereas, in the case of a fixed Pauli axis, we only have errors of one type (either $X, Y$ or $Z$). For a fixed error budget, specified by fidelity, the case of a non-Pauli axis results in the error strength spread over a larger number of correctable errors than the case of a fixed Pauli axis which would include relatively higher weight Pauli errors of one type. Hence, the Steane code has better error correction capability. Figure \ref{fig:arbitRotHaarPseduoThresh} provides evidence to our argument by showing that the threshold angle for performance gains from RC under rotations about various axes, is higher for non-Pauli axes compares to the Pauli ones. As a consequence, we also note that for rotation angles $\omega_{\star} < \omega < \ol{\omega}_{\star}$, the largest gains from RC are achieved for rotations axes that lie between the $X, Y$ and $Z$ axes as opposed to the individual Pauli axes.

\subsection{Composition of coherent and stochastic map}
\begin{figure}[h]
\centering
\includegraphics[scale=0.65]{\figsdir/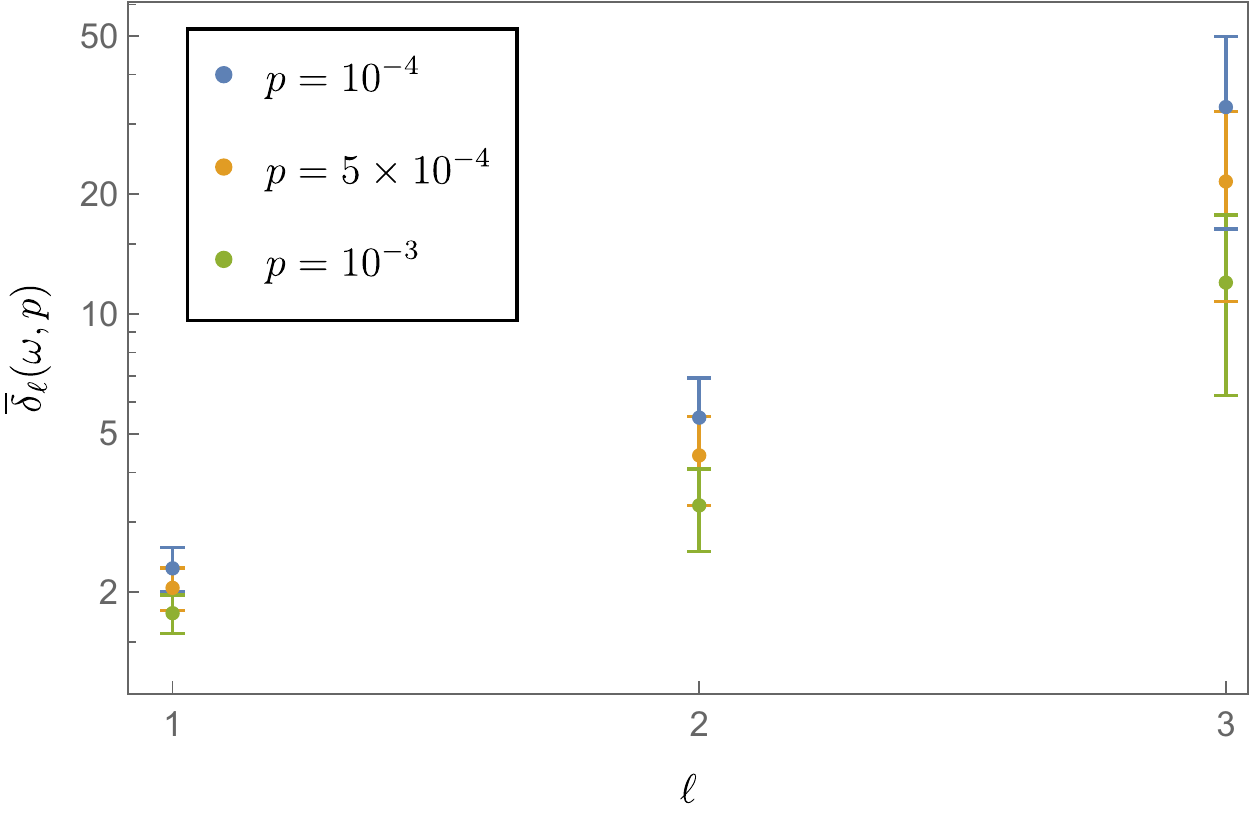}
\caption{The impact of the depolarizing component on the gains from RC. We fix the average infidelity per qubit to be $r \approx 0.003$ and increase the value of the depolarizing strength from $p = 10^{-4}$ to $p= 10^{-3}$. The value of $\omega$ corresponding to each value of $p$ is chosen such that the total physical infidelity of the qubit remains constant. We observe that the gains from RC diminish with increase in depolarizing strength. This is because RC does not impact the stochastic component of the noise model.}
\label{fig:genericNoiseHaar}
\end{figure}

So far, we have shown that RC always improves the performance of quantum error correcting codes under coherent errors. Generic unital maps can be approximately described as a composition of a coherent error and a Pauli error model \cite{CME19,CWE19}. In what follows, we consider a more general unital map where we model coherent errors in a similar fashion as in the previous section and for Pauli errors, we choose the depolarizing error model, i.e., 
\begin{equation}
\cE \simeq (\cE_{\textit{dep}} \circ \cE_{\textit{coh}})^{\otimes n},
\label{eq:ErrorModel}
\end{equation}
where 
\begin{align}
\cE_{\textit{coh}}(\rho) &= U \rho U^\dagger,\nonumber\\
\cE_{\textit{dep}}(\rho) &= (1-p) \rho + \frac{p}{2} \mathbb{I}.
\end{align}
and $U$ can be parameterized using Eq.~\eqref{eq:arbitUnitary}. In what follows, we will study the impact of RC under the approximation given by Eq.~\eqref{eq:ErrorModel}. Note that both the coherent as well as the incoherent parts of the error model contribute to the strength of noise, for instance, the average gate fidelity. While RC only affects the coherent part of the error process, we expect that for a fixed noise strength, the performance gain due to RC under the error model described above will diminish with increasing $p$. This expectation is supported by the numerical simulations presented in Fig.~\ref{fig:genericNoiseHaar}, where we present numerical estimates of $\ol{\delta}_{\ell}(\omega,p)$ for several depolarizing strengths $p$.  Here, $\ol{\delta}_{\ell}(\omega,p)$ is defined analogous to Eq.~\eqref{eq:ave_gain_arbitrary_rot} as
\begin{gather}
\ol{\delta}_{\ell}(\omega,p) = \frac{1}{2 \pi} \int_{0}^{2\pi} \int_{0}^{\pi} \ths \delta_{\ell}(\theta, \phi, \omega, p) \ths \sin(\theta) \ths d\theta \ths d\phi \ths . \label{eq:ave_gain_arbitrary_rotp}
\end{gather}

Note that in all of the error models considered so far, we have only observed gains in performance due to RC. However, amongst the most general CPTP maps including the unital as well as non-unital types, we have identified cases under which RC can lead to a loss in the performance. Some examples of these maps are mentioned in Appendix~\ref{sec:app_numerical_results}.

\section{Conclusion}
\label{sec:conclusion}
The application of randomized compiling in fault tolerance is attractive for two reasons. First, amongst the exponentially growing number of parameters controlling a physical noise process, RC effectively eliminates the impact of most of them on a QEC scheme. Second, since RC removes multiple noise sources, we expect the code to perform better. This paper provides concrete evidence to show that RC improves the performance of quantum error correction under a wide class of coherent errors. We have identified noise regimes where gains are drastic for the case of concatenated Steane codes. In particular, it grows doubly exponentially with the number of levels, under small rotations about a Pauli axis. Our results can be extended to guarantee performance gains under generic unital noise processes, leveraging tools from \cite{CME19,CWE19} that approximate a unital noise process as a composition of a coherent and an incoherent error model. These observations strengthen the need for active noise tailoring methods as a crucial component of a fault tolerant scheme. 

Performance gains offered by RC also depend on the strength of errors affecting the physical qubits. We stumbled upon an interesting observation that indicates gains decrease when the amount of coherent rotation error passes beyond a threshold value. To the best of our knowledge a threshold of this nature hasn't been reported in earlier works. The threshold helps estimate the maximum possible noise that can be alleviated on a hardware device by leveraging RC tools. We also carried out extensive studies to analyze the variation of this threshold with the features of the underlying coherent error model.

Beyond the paradigm of identical unital maps across all physical qubits, we argue that unilateral conclusions about performance gains due to RC cannot be made, i.e., it depends strongly on the microscopic details of the underlying physical noise process. Our arguments are strengthened by numerical studies of complex physical noise processes that revealed some cases where the code's performance can also degrade in the presence of RC. In Ref.~\cite{CWBL17}, it was shown that twirled noise processes may improve or degrade thresholds depending on the decoding algorithm used. In this paper we arrive at a similar conclusion by exploring different error models for the minimum weight decoder.

Obtaining efficiently computable estimates for performance gains due to RC in different experimental setups would be crucial to optimizing fault tolerance resources in near-term applications. In the absence of exact values, it would be useful to provide bounds for the impact of RC on the code’s performance. Although RC's impact on performance depends strongly on the underlying noise process, it is still interesting to see that it can provide significant gains for a wide variety of realistic error models and relevant error regimes. 

To ensure a performance gain from a noise tailoring technique, such as RC, ideally, we want to cancel the impact of those terms in the underlying noise process, which correspond to uncorrectable errors -- since these add to the logical infidelity. It would be worthwhile to explore ways of controlling physical noise sources to ensure that RC always offers a gain in performance. It would also be interesting to explore different Twirling gate sets that can tailor the noise process to suppress terms that contribute negatively to the logical channel’s fidelity. Although we identified a handful of cases where a performance loss is observed, it will be noteworthy to develop cheap experimental protocols to ascertain whether performing error correction with RC will be significantly beneficial for a given device.

\begin{acknowledgments}
This research was undertaken thanks in part to funding from the Canada First Research Excellence Fund. Research was partially sponsored by the ARO and was accomplished under Grant Number: W911NF-21-1-0007.  SDB acknowledges support from the Australian Research Council (ARC) via the Centre of Excellence in Engineered Quantum Systems (EQuS) project number CE170100009.  
\end{acknowledgments}

\bibliography{refs}

\begin{appendix}
\begin{widetext}
\section{Logical fidelity calculation for rotation about Z-axis}
\label{sec:app_log_fid}
In this appendix section, we will derive the logical performance of Steane code under a unitary noise process described by a small over-rotation about the $Z-$axis, i.e. $\cE(\rho) = R_{Z}(\omega)\rho R_{Z}(-\omega)$ where
\begin{gather}
R_{Z}(\omega) = \cos (\omega/2) \ths I + i \sin (\omega/2) \ths Z \ths . \label{eq:rtz2}
\end{gather}

Recall that the Steane code is a $[[n,k]]$ quantum code with $n = 7, k = 1$, whose encoded states are fixed by the Stabilizer group $\cS$ generated by $n-k$ generators:
\begin{gather}
\cS = \langle ZZZZIII, ZZIIZZI, ZIZIZIZ, XXXXIII, XXIIXXI, XIXIXIX\rangle \ths . \label{eq:steane_gens}
\end{gather}
The effect of the unitary noise in eq. \ref{eq:rtz} on each of the $n$ qubits in the encoded state can be written as
\begin{flalign}
\cE^{\ot n}(\bar{\rho}) &= R^{\ot n}_{Z}(\omega) \ths \bar{\rho} \ths R^{\ot n}_{Z}(-\omega) \nonumber \\
& = \sum_{w \ts\in\ts \bZ^{2n}_{2}} (-1)^{\sum_{j=n+1}^{2n} w_j} (\cos (\omega/2))^{2n-|w|} (i \sin (\omega/2))^{|w|} \left(\ot_{j=1}^{n} Z^{w_j}\right) \bar{\rho} \left(\ot_{j=n+1}^{2n} Z^{w_j}\right) \label{eq:enc_rtz2} \ths .
\end{flalign}
where $|w|$ is the Hamming weight of the binary sequence $w \in \bZ^{2n}_{2}$.

To understand the effect of RC on performance, we need to estimate the total contribution to logical fidelity from terms in the noise process whose effect is rendered useless by RC. Since the noise model in eq. \ref{eq:enc_rtz2} only applies $Z-$type errors, it suffices to consider the effect of correctable errors $E$ and $E^{\prime}$ that are purely $Z-$type, besides the identity. In other words, $E, E^{\prime} \in \langle Z_{1}, Z_{2}, \ldots, Z_{n}\rangle$. Table \ref{tab:offdiag} shows the contribution to the logical fidelity that is eliminated by RC. Each of the four rows in the table is associated with a $\chi-$matrix element of a particular form, labelled by $\gamma_{i}$ for $1\leq i \leq 4$.

\begin{table}[h]
\begin{center}
\begin{subtable}{0.9\textwidth}
\begin{tabular}{| c | c | c | c  | c |} 
\hline
$E$ & $E^{\prime}$ & Condition on $E$ and $E^\prime$ & $\chi_{E,E^{\prime}}$ & \\
\hline
$I^{\otimes 7}$ & $S$ & $S \in \cS\ts\backslash\ts \{\bI\}$ & $\cos^{10} (\omega/2) \sin^4 (\omega/2)$ & = $\gamma_{1}$ \\
\hline
$Z_i$ & $Z_i S$ & $S \in \cS\ts\backslash\ts \{\bI\}$ , $1 \leq i \leq 7$ & $\cos^8 (\omega/2) \sin^4 (\omega/2) \ths (3 \sin^{2} (\omega/2) - 4 \cos^{2} (\omega/2))$ & = $\gamma_{2}$\\
\hline
$S$ & $S^{\prime}$ & $S, S^{\prime} \in \cS\ts\backslash\ts \{\bI\}$ , $S \neq S^{\prime}$ & $\cos^6 (\omega/2) \sin^8 (\omega/2)$ & = $\gamma_{3}$ \\
\hline
$Z_i S$ & $Z_i S^{\prime}$ & $S,S^{\prime} \in \cS\ts\backslash\ts \{\bI\}$ , $S \neq S^{\prime}$ , $1 \le i \le 7$& \makecell{$6 \cos^8 (\omega/2) \sin^6 (\omega/2) -12 \cos^6 (\omega/2) \sin^8 (\omega/2)$ \\ $+ 3 \cos^4 (\omega/2) \sin^{10} (\omega/2)$} & = $\gamma_{4}$ \\
\hline
\end{tabular}
\caption{}
\label{tab:offdiag}
\end{subtable}
\begin{subtable}{0.9\textwidth}
\begin{tabular}{| c | c | c | c  | c |} 
\hline
$E$ & $E^{\prime}$ & Condition on $E$ and $E^\prime$ & $\chi_{E,E^{\prime}}$ & \\
\hline
$I^{\otimes 7}$ & $I^{\otimes 7}$ &  & $\cos^{14} (\omega/2)$ & = $\kappa_1$ \\
\hline
$S$ & $S$ & $S \in \cS\ts\backslash\ts \{\bI\}$ & $\cos^6 (\omega/2) \sin^8 (\omega/2) $ & = $\kappa_2$ \\
\hline
$Z_i$ & $Z_i$ & $1 \le i \le 7$ & $\cos^{12} (\omega/2) \sin^2 (\omega/2)$ & = $\kappa_3$ \\
\hline
$Z_i S$ & $Z_i S$ & $S \in \cS\ts\backslash\ts \{\bI\}$, $1 \le i \le 7$& $4 \cos^8 (\omega/2) \sin^6 (\omega/2) + 3 \cos^4 (\omega/2) \sin^{10} (\omega/2)$ & = $\kappa_4$ \\
\hline
\end{tabular}
\caption{}
\label{tab:diag}
\end{subtable}
\caption{The above table describes the contribution to logical fidelity from different types of elements of the physical channel. While table \ref{tab:diag} describes the contribution to the logical infidelity from the diagonal (Pauli) terms, table \ref{tab:offdiag} specifies that from the off-diagonal terms in the physical channel. In each of the tables, the total contribution to logical infidelity is divided into four categories: (i) labelled $\gamma_{1}, \gamma_{2}, \gamma_{3}$ and $\gamma_{4}$ for the off-diagonal terms and (ii) $\kappa_{1}, \kappa_{2}, \kappa_{3}$ and $\kappa_{4}$ for the diagonal terms.}
\label{tab:loginfid_contributions}
\end{center}
\end{table}

Table \ref{tab:diag} provides all the ingredients necessary to compute the logical infidelity of the Steane code under the RC setting:
\begin{flalign}
r(\ol{\cE^T}_{1}) & = 1 - (\kappa_{1} + 7 \kappa_{2} + 7 \kappa_{3} + 7 \kappa_{4}) \label{eq:r_RC_with_multiplicities} \ths , \\
& = \frac{1}{512} (256 - 231 \cos(\omega) - 49 \cos(3 \omega) + 21 \cos(5 \omega) + 3 \cos(7 \omega)) \ths . \label{eq:app_loginfid_RC}
\end{flalign}
Note that the coefficient appearing alongside each $\phi_{i}$ in eq. \ref{eq:r_RC_with_multiplicities} corresponds to its multiplicity, i.e., the number of combinations of errors $E, E^{\prime}$ that result in the same value of $\phi_{i}$. In the absence of RC, the logical infidelity can be calculated using both tables \ref{tab:offdiag} and \ref{tab:diag}:
\begin{flalign}
r(\ol{\cE}_{1}) &= 1-(\kappa_{1} + 7 \kappa_{2} + 7 \kappa_{3} + 7 \kappa_{4} + 14 \gamma_{1} + 14 \gamma_{2} + 42 \gamma_{3} + 14 \gamma_{4}) \ths ,\nonumber \\
&= \frac{1}{64} (32 - 21 \cos (\omega)  - 14 \cos (3 \omega ) + 3 \cos (7 \omega )) \ths . \label{eq:final_impact_RC}
\end{flalign}

The above expressions describe the logical infidelities for level$-1$ concatenated Steane code in the RC and non-RC settings. The gain $\delta_1$ can be calculated as the ratio of the above quantities. The appendix section \ref{sec:app_chilogapp} discusses the recursion to compute the average logical channel for level$-\ell$ concatenated Steane code followed by the computation of the different metrics at level$-\ell$.
\section{Logical channel for the concatenated Steane code} \label{sec:app_chilogapp}
In this appendix section, we will describe the computation of the average logical channel for the level$-\ell$ concatenated Steane code under rotations about the $Z-$axis described in section \ref{sec:results}. Ideally, we would like to take an exact average over conditional channels corresponding to all possible syndromes of the level$-\ell$ concatenated Steane code. However, the number of syndromes and hence the number of conditional channels grow exponentially with the number of physical qubits and the analysis becomes intractable beyond a few levels. Instead, in this section we compute an approximation wherein we recurse over the individual entries of the level$-1$ logical channel to arrive at the level$-\ell$ logical channel. We will achieve this in two broad steps:
\begin{enumerate}
	\item Computation of level$-1$ logical channel.
	\item Establish a recursion to compute level$-(\ell+1)$ from level$-\ell$ logical channel.
\end{enumerate}

For a given noise process $\cE$, we refer to its $\chi-$matrix as $\chi(\cE)$ and the corresponding logical $\chi-$matrix as $\chi(\ol{\cE}_1)$. The following equation prescribes a way to calculate the entries of $\chi(\ol{\cE})$ from ${\chi}(\cE)$ \cite{IP17}.

\begin{gather}
\chi(\ol{\cE}_{1})_{l,m} = \sum_{\substack{E,E^{\prime} \in \cE_{C} \\ s(E) = s(E^{\prime}) \ts , \ts \ol{E} = \ol{E}^{\prime}}} \phi(E, l) \ths  \phi^{\star}(E^\prime, m) \ths \chi_{E \ol{P}_{l},\ol{P}_{m}E^{\prime}} \ths .\label{eq:chi_lm_log_phys_app}
\end{gather}
where $\cE_{C}$ refers to the set of correctable errors, $\ol{P}_{i}$ denotes the logical version of Pauli $P_i$, and $R_{s(E)} \ths |E \ths \ol{P}_{l}| = \phi(E,l) \ths S \ths |\ol{P}_{l}|$, for $l \in \{0,1,2,3\}$, any Pauli error $E$ and some stabilizer $S$. Here $|P|$ stands for the bare Pauli without any associated global phase. Note that, since the error model is a rotation about the $Z-$axis, we have $\cE_{C} = \langle \{S_j Z_i \ts : \ts 1 \le i \le n \ts , \ts S_{j} \in \cS_{Z}\}\rangle$. Here $Z_i$ refers to a single qubit $Z$ error on qubit $i$ and $\cS_{Z} = \langle ZZZZIII, ZZIIZZI, ZIZIZIZ \rangle$. 

It is easy to see that the average logical channel for the level$-\ell$ concatenated Steane code $\chi(\ol{\cE}_{\ell})$ takes the form \cite{HDF18}:
\begin{equation}
	\chi(\ol{\cE}_{\ell}) = \begin{pmatrix}
		& [\chi(\ol{\cE}_{\ell})]_{0,0} & 0 & 0 & [\chi(\ol{\cE}_{\ell})]_{0,3}\\
		& 0 & 0 & 0 & 0\\
		& 0 & 0 & 0 & 0\\
		& ([\chi(\ol{\cE}_{\ell})]_{0,3})^{*} & 0 & 0 & 1-[\chi(\ol{\cE}_{\ell})]_{0,0}
	\end{pmatrix},
\label{eq:chiLevelLapp}
\end{equation}
where $([\chi(\ol{\cE}_{\ell})]_{0,3})^{*}$ denotes the complex conjugate of $[\chi(\ol{\cE}_{\ell})]_{0,3}$. 

First, we compute the entries for the level$-1$ matrix $\chi(\ol{\cE}_{1})$. Using Table \ref{tab:offdiag}, we have
\begin{flalign}
[\chi(\ol{\cE}_{1})]_{0,0} & = \kappa_1 + 7 \sum_{i=1}^{3} \phi_i + 28 \ths \chi_3 + 14 \sum_{j=1}^{4} \chi_j \ths , \nonumber \\
& = \frac{1}{64} (21 \cos (\omega )+14 \cos (3 \omega )-3 \cos (7 \omega )+32) \ths . \label{eq:chi00app}
\end{flalign}

\begin{table}[h]
	\begin{center}
			\begin{tabular}{| c | c | c | c  | c |} 
				\hline
				$E$ & $\ol{Z}E^{\prime}$ & Condition on $E$ and $E^\prime$ & $\chi_{E,\ol{Z}E^{\prime}}$ & \\
				\hline
				$I^{\otimes 7}$ & $\ol{Z}$ &  & $i \sin ^7(\omega/2) \cos ^7(\omega/2)$ & = $\zeta_1$  \\
				\hline
				$I^{\otimes 7}$ & $\ol{Z}S$ & $S \in \cS_Z\ts\backslash\ts \{\bI\}$ & $i \sin ^3(\omega/2) \cos ^{11}(\omega/2)$ & = $\zeta_2$  \\
				\hline
				$S$ & $\ol{Z}$ & $S \in \cS_Z\ts\backslash\ts \{\bI\}$ & $i \sin ^{11}(\omega/2) \cos ^3(\omega/2)$ & = $\zeta_3$ \\			
				\hline
				$S$ & $\ol{Z}S'$ & $S,S^{\prime} \in \cS_Z\ts\backslash\ts \{\bI\}$ , $S \neq S^{\prime}$ & $i \sin ^7(\omega/2) \cos ^7(\omega/2)$  & = $\zeta_4$\\
				\hline
				$Z_i$ & $\ol{Z}Z_i$ & $1 \le i \le 7$ & $-i \sin ^7(\omega/2) \cos ^7(\omega/2)$ & = $\zeta_5$\\
				\hline
				$Z_i$ & $\ol{Z}Z_iS$ & $S \in \cS_Z\ts\backslash\ts \{\bI\}$, $1 \le i \le 7$  & $4 i \sin ^5(\omega/2) \cos ^9(\omega/2)-3 i \sin ^3(\omega/2) \cos ^{11}(\omega/2)$ &= $\zeta_6$\\
				\hline
				$Z_iS$ & $\ol{Z}Z_i$ & $S \in \cS_Z\ts\backslash\ts \{\bI\}$, $1 \le i \le 7$ & $4 i \sin ^9(\omega/2) \cos ^5(\omega/2)-3 i \sin ^{11}(\omega/2) \cos ^3(\omega/2)$ &= $\zeta_7$\\
				\hline
				$Z_iS$ & $\ol{Z}Z_iS'$ &$S,S^{\prime} \in \cS_Z\ts\backslash\ts \{\bI\}$ , $S \neq S^{\prime}$, $1 \le i \le 7$  & \makecell{$12 i \sin ^5(\omega/2) \cos ^9(\omega/2)-25 i \sin ^7(\omega/2) \cos ^7(\omega/2)$ \\ $+12 i \sin ^9(\omega/2) \cos ^5(\omega/2)$}& = $\zeta_8$\\
				\hline
				\end{tabular}
		\caption{The above table describes the contribution to $\ol{\chi}_{0,3}(\cE)$ from different types of elements of the physical channel. Note that none of these contributions come from the diagonal part of $\chi(\cE)$.}
		\label{tab:chi03}
	\end{center}
\end{table}

Table \ref{tab:chi03} provides all the ingredients necessary to compute $[\chi(\ol{\cE}_{1})]_{0,3}$. Taking into account the multiplicities of terms of each kind, we have
\begin{flalign}
  [\chi(\ol{\cE}_{1})]_{0,3} & = \zeta_1 + 42 \ths \zeta_4 + 7 \sum_{i=2}^{8} \zeta_i \ths , \\
 & = -\frac{1}{8} i \sin ^3(\omega)  (9 \cos (2 \omega )+3 \cos (4 \omega )+2) \ths .
 		\label{eq:chi03app}
\end{flalign}

In the second step, we establish a recursion to compute the individual entries of $\chi(\ol{\cE}_{\ell})$ from the entries of $\ol{\chi}_{\ell-1}(\cE)$ under hard-decoding algorithm. After massaging the expressions in equations \ref{eq:chi00app} and \ref{eq:chi03app}, we observe that 
\begin{flalign}
	[\chi(\ol{\cE}_{\ell+1})]_{0,0} & = f_{0,0}([\chi(\ol{\cE}_{\ell})]_{0,0}), \text{and}\\
[\chi(\ol{\cE}_{\ell+1})]_{0,3} & = f_{0,3}([\chi(\ol{\cE}_{\ell})]_{0,3}),
\end{flalign}
where
\begin{flalign}
f_{0,0}(z) & = z^2 (63 - 434 z + 1260 z^2 - 1848 z^3 + 1344 z^4 - 384 z^5), \text{and}\nonumber\\
f_{0,3}(z) & = -2 z^3 (7 + 84 z^2 + 192 z^4).
\label{eq:recursion}
\end{flalign}
Combining the above two steps, we compute all the entries of $[\chi(\ol{\cE}_{\ell+1})]$. 

For small rotation angle $\omega$, we observe from equations \ref{eq:chi00app} and \ref{eq:recursion} that upto leading order 
\begin{flalign}
	[\chi(\ol{\cE}_{\ell})]_{0,0} & \approx 1 - 63^{2^\ell - 1} \ths (\omega/2)^{2^{\ell+1}}, \text{and} \nonumber\\
		[\chi(\ol{\cE}_{\ell})]_{0,3} & \approx - i 14^{\frac{3^\ell - 1}{2}} \ths (\omega/2)^{3^{\ell}}.
\end{flalign}
Note that with increase in number of levels $\ell$, for small angle $\omega$, $	[\chi(\ol{\cE}_{\ell})]_{0,0} \rightarrow 1$ and $	[\chi(\ol{\cE}_{\ell})]_{0,3} \rightarrow 0$. This is expected because for small angles, the channel is close to the identity channel and the error correction procedures is able to correct all the errors. Also, note that the off diagonal entry approaches $0$ faster than the diagonal entry approaches $1$. This is a consequence of the process of error correction decohering the physical channel \cite{BWGB18}. 

Now, we compute the logical $\chi-$matrix corresponding to the noise process under RC i.e. $\chi(\ol{\cE^T}_{\ell})$. The matrix in this case takes the form:
\begin{gather}
	\chi(\ol{\cE^T}_{\ell}) = \begin{pmatrix}
		& [\chi(\ol{\cE^T}_{\ell})]_{0,0} & 0 & 0 & 0\\
		& 0 & 0 & 0 & 0\\
		& 0 & 0 & 0 & 0\\
		& 0 & 0 & 0 & 1-[\chi(\ol{\cE^T}_{\ell})]_{0,0}
	\end{pmatrix}.
	\label{eq:chiRCLevelLapp}
\end{gather}
Similar to the nonRC case, we first compute the entries for the level$-1$ matrix $\chi(\ol{\cE^T}_{1})$. Using the ingredients from table \ref{tab:diag}, we have
\begin{gather}
[\chi(\ol{\cE^T}_{1})]_{0,0} = \frac{1}{512} (256 + 231 \cos(\omega) + 49 \cos(3 \omega) - 21 \cos(5 \omega) - 3 \cos(7 \omega)) \ths . \nonumber
\end{gather}
The recursive relation to calculate the above quantity for higher levels is given by:
\begin{gather}
[\chi(\ol{\cE^T}_{\ell+1})]_{0,0} = g_{0,0}([\chi(\ol{\cE^T}_{\ell})]_{0,0}) \ths , \nonumber
\end{gather}
where
\begin{gather}
g_{0,0}(z) = z^2 (21 - 98 z + 210 z^2 - 252 z^3 + 168 z^4 - 48 z^5) \ths . \nonumber
\end{gather}
For small rotation angle $\omega$, upto leading order
\begin{gather}
[\chi(\ol{\cE^T}_{\ell})]_{0,0} \approx 1 - 21^{2^\ell - 1} \ths (\omega/2)^{2^{\ell+1}} \ths . \nonumber
\end{gather}
The above expression indicates that $[\chi(\ol{\cE^T}_{\ell})]_{0,0} \rightarrow 1$ with increase in number of concatenation levels $\ell$ provided the angle of rotation is below the threshold.

Having arrived at an expression for the average logical channel for a level$-\ell$ concatenated code, we can now define the logical error rate using the infidelity and diamond distance metrics. The logical infidelity takes the simple closed form:
\begin{flalign}
\ol{r}_{\ell} = 1 - [\chi(\ol{\cE}_{\ell})]_{0,0} \nonumber \ths .
\end{flalign}

\section{Numerical results for complex noise models} \label{sec:app_numerical_results}
In this appendix section, we will present numerical studies of the performance of concatenated Steane codes under two distinct models of general Markovian noise. The results are presented as scatter plots formatted as follows. Each point is associated to the performance of a physical noise process. While the $X-$ coordinate is used to denote the physical error rate, its $Y-$coordinate denotes the ratio between the performance in the non-RC setting and the RC setting, measured by $\delta_{\ell}$ in eq. \ref{eq:deltaDef}. Note that RC can either improve or degrade the code's performance. We have used a dashed line at $\delta_{\ell} = 1$ to identify the breakeven region where RC has no impact on the performance. Points that lie below the dashed line, coloured in red, identify physical channels where a degradation in performance is observed. On the other hand, points in green that lie above the dashed line identify physical channels where RC provides a performance gain. The points in grey, that lie close to the dashed line should be ignored since they correspond to cases where the relative difference between the logical error rates for the non-RC and RC cases is negligible: less than $10\%$.

The first complex error model is a unitary model where each qubit experiences a different random rotation about an arbitrary non-Pauli axis $\hat{n}$, specified by $U$ of the form
\begin{gather}
U = e^{-i \frac{\pi}{2}\delta \hat{n}\cdot \vec{\sigma}} , \label{eq:qubit_unitary_error}
\end{gather}
where $\delta$ is the angle of rotation. Hence, the $n-$qubit unitary errors in our model are of the form $\otimes_{i=1}^{n} U_i$, where $U_{i}$ in prescribed by eq. \ref{eq:qubit_unitary_error}. We control the noise strength by setting the rotation angles $\delta_i$ drawn from the normal distribution: $\cN(\mu_\delta, \mu_\delta)$, where $10^{-3}\leq \mu_\delta\leq 10^{-1}$. Fig. \ref{fig:rtas} shows the performance gain metric under this error model. It demonstrates that there exist some instances where RC provides a performance gain of $10$x, as well as others where RC causes a performance degradation of $10$x.

\begin{figure}[h]
\begin{subfigure}[t]{0.9\textwidth}
\centering
\includegraphics[scale=\predictscale]{\figsdir/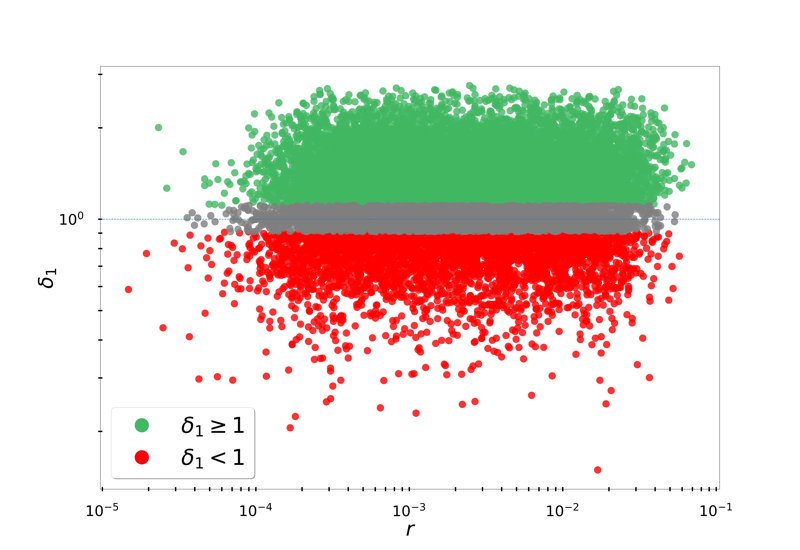}
\caption{}
\label{fig:pg_1cplot_rtas_infid_infid}
\end{subfigure}%

\vspace{\predvspace}

\begin{subfigure}[t]{0.9\textwidth}
\centering
\includegraphics[scale=\predictscale]{\figsdir/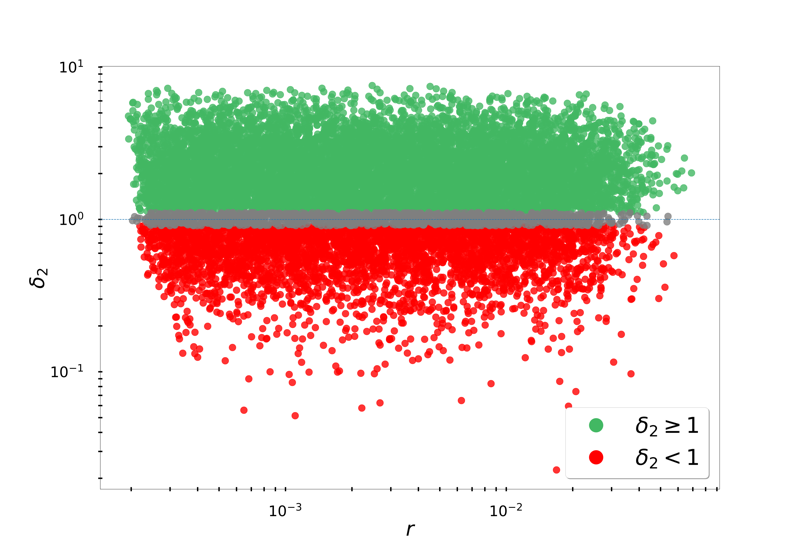}
\caption{}
\label{fig:pg_2cplot_rtas_infid_infid}
\end{subfigure}
\caption{The above figures highlight the strong dependence of the impact of RC on the details of the physical noise process, for concatenated Steane codes. The ensemble of noise processes considered here comprises of $16000$ samples of unitary rotations about a fixed random axis. Red and green points are used to identify physical noise processes that lead to a performance gain and a performance loss, respectively, in the presence of RC. The magnitude of performance gains and losses are measured by the ratio of logical error rates in the non-RC and RC settings, i.e., $\delta_{1}$ for level-1 concatenated Steane code in fig. \ref{fig:pg_1cplot_rtas_infid_infid}, and $\delta_{2}$ for level-2 in fig. \ref{fig:pg_2cplot_rtas_infid_infid}.} 
\label{fig:rtas}
\end{figure}

The second error model is described by the i.i.d action of a random single qubit CPTP map, on each of the physical qubits of the code. The random CPTP map on a single qubit is derived from unitary dynamics $U$ on a Hilbert space of three qubits \cite{IP17}. The unitary matrix $U$ is generated form a random Hermitian matrix $H$ using $U = e^{-i H t}$, where $0 \leq t \leq 1$ provides a handle on the strength of noise described by the resulting CPTP map. We vary the noise strength by controlling $t$ in the range $[0.001, 0.1]$. Figure \ref{fig:cptp} shows RC’s impact on the performance of concatenated Steane codes under physical CPTP maps. The absence of a clear trend showing a performance gain or degradation is evident for level-2 in fig. \ref{fig:pg_2cplot_cptp_infid_infid}. Even across physical CPTP maps with similar fidelity, while for one instance, RC induces a performance gain of up to three orders of magnitude, for another, it inflicts a loss in performance of similar magnitude.

\begin{figure}[h]
\begin{subfigure}[t]{0.9\textwidth}
\centering
\includegraphics[scale=\predictscale]{\figsdir/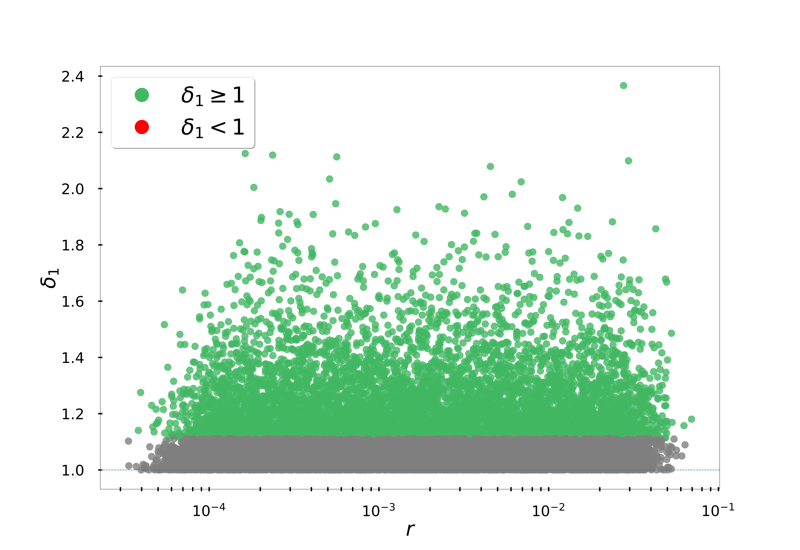}
\caption{}
\label{fig:pg_1cplot_cptp_infid_infid}
\end{subfigure}%

\vspace{\predvspace}

\begin{subfigure}[t]{0.9\textwidth}
\centering
\includegraphics[scale=\predictscale]{\figsdir/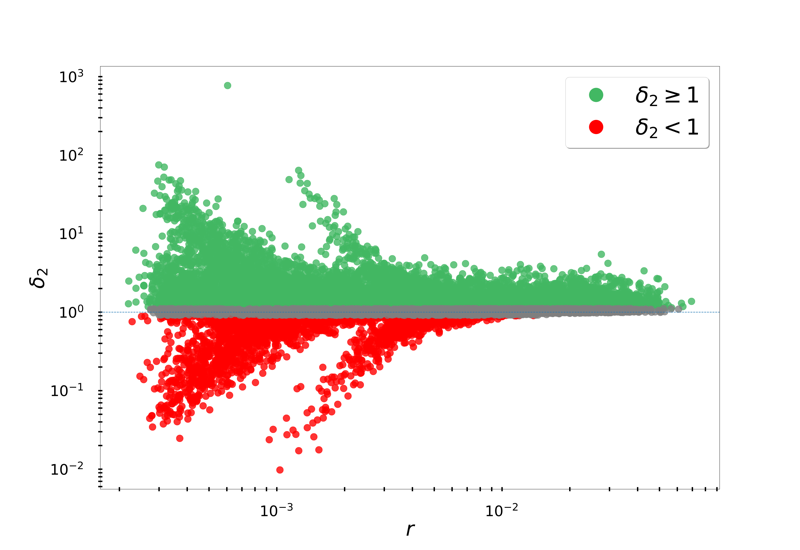}
\caption{}
\label{fig:pg_2cplot_cptp_infid_infid}
\end{subfigure}
\caption{The above figures highlight the strong dependence of the impact of RC on the details of the physical noise process, for concatenated Steane codes. The ensemble of noise processes considered here comprises of $18000$ random CPTP maps. Red and green points are used to identify physical noise processes that lead to a performance gain and a performance loss, respectively, in the presence of RC. The magnitude of performance gains and losses are measured by the ratio of logical error rates in the non-RC and RC settings, i.e., $\delta_{1}$ for level-1 concatenated Steane code in fig. \ref{fig:pg_1cplot_cptp_infid_infid}, and $\delta_{2}$ for level-2 in fig. \ref{fig:pg_2cplot_cptp_infid_infid}.} 
\label{fig:cptp}
\end{figure}

The case of level-1 performance under physical CPTP maps in fig. \ref{fig:pg_1cplot_cptp_infid_infid} is rather different from the level-2 case in \ref{fig:pg_2cplot_cptp_infid_infid}. Over the large ensemble of $18000$ physical CPTP maps, we observe that RC always leads to performance gains for the level-1 Steane code. These performance gains can be explained as follows. First of all, a CPTP map can be well approximated by its leading Kraus operator $K$, which is derived from the largest eigenvector of its Choi matrix \cite{CME19}. Furthermore, in an i.i.d physical error model, $K$ can be expressed as a tensor product. In terms of $K$, the leading contributions to infidelity come from chi-matrix entries $\chi_{i,j}$ expressed as:
\begin{gather}
\chi_{i,j} = \Tr(K P_{i}) \Tr(K^{\dagger} P_{j}) \ths , \label{eq:chi_kraus}
\end{gather}
where $P_{i}$ is a single qubit of one type (X, Y or Z), and $P_{j} = P_{i} S$ for some stabilizer $S$, is a three-qubit error of the same type as $P_{i}$. In the low noise regime, the off-diagonal entries of $K$ are small, especially for incoherent CPTP maps, where $K$ is close to a Positive sem-definite matrix \cite{CME19}. Using the fact that the trace inner product between $K$ and the Pauli matrix $Z$ is a real number $d$ given by $d = K_{1,1} - K_{2,2}$, we can conclude that $\chi_{i,j}$ in eq. \ref{eq:chi_kraus} for $Z-$type errors $P_{i}$ and $P_{j}$ of weights 1 and 3 respectively, is always positive. In other words, the $\chi_{i,j} \sim d^{4}$ for some $d \ll 1$. Removing such terms should degrade the performance of the code. On the contrary, removal of $\chi_{i,j}$ for uncorrectable errors $P_{i}, P_{j}$ leads to performance gains. The largest of these chi-matrix entries can be identified with two $Z-$type Pauli errors $P_{i}, P_{j}$, each having weight two. This property can be associated with the fact that the Steane code is degenerate: there exists a logical operator whose weight is smaller than that of a stabilizer. Repeating a similar analysis as before, we find that the corresponding $\chi_{i,j}$ for these uncorrectable errors, also scale as $d^{4}$ for some $d \ll 1$. Their removal leads to performance gains. Note that there are more uncorrectable errors than correctable ones and the corresponding chi-matrix elements have comparable magnitudes. Hence, we find that RC is more likely to induce performance gains. Note that higher concatenation levels of the Steane code do not correspond to degenerate codes. Hence, we cannot guarantee a performance gain or degradation in those cases, as shown in fig. \ref{fig:pg_2cplot_cptp_infid_infid}

\section{Applying RC in a Fault tolerance scheme} \label{sec:app_qec_rc}

Randomized compiling \cite{WE16} is a technique that tailors a general Markovian noise process into an effective Pauli noise process. In this appendix section, we will briefly review how we can apply randomized compiling in a fault-tolerant setting. This procedure was first described in Ref.~\cite{Iyer2021}.

In fault tolerant circuits, each logical gate $\ol{G}$ is sandwiched between quantum error correction ($\QEC$) routines. Similar to Ref.~\cite{WE16}, we divide logical gates into two sets: $\cS_{1}$ and $\cS_{2}$, calling them easy and hard gates respectively. A requirement for $\cS_{1}$ and $\cS_{2}$ is
\begin{gather}
	\ol{G}~T~\ol{G}^{\dagger}~\QEC = \QEC(T)~\ol{C} \ths \label{eq:P_map_easy}
\end{gather}
for all easy logical gates $\ol{C}\in\cS_{1}$, $n-$qubit Pauli gates $T$ and hard gates $\ol{G}$. Here $\QEC(T)$ refers to the compilation of the Pauli gate $T$ in the $\QEC$ routine. The above requirement can be proven to be true using results from standard randomized compiling \cite{Iyer2021}. 

Figure \ref{fig:clock_cycle} shows a canonical presentation of a quantum circuit, where the $k$-th clock cycle is composed of an easy gate $\ol{C}_{k}$ and a hard gate $\ol{G}_{k}$, sandwiched between $\QEC$ routines. Noise processes affecting easy and hard gates are denoted by $\cE_{1,k}$ and $\cE_{2,k}$ respectively. These complex processes can be tailored to Pauli errors by inserting Pauli gates $T_{1,k}, T^{\dag}_{1,k},T_{2,k}, T^{\dag}_{2,k}$ as shown in fig. \ref{fig:insert_twirling}. However, to guarantee that they be applied in a noiseless fashion, we compile them into the existing gates in the fault tolerant circuit. This is achieved as follows. First $T^{\dagger}_{1,k}$ and $T_{2,k}$ are compiled into $\QEC$ following $\cE_{1,k}$, resulting in $\QEC(T^{\dagger}_{1,k}T_{2,k})$. Secondly, $T^{\dagger}_{2,k}$ is propagated across $\ol{G}_{k}$, and compiled with $\QEC~\ol{C}_{k+1}T_{k+1}$, resulting in a \emph{dressed gate} $\ol{C^{D}_{k+1}}$ defined by
	\begin{gather}
		\ol{C^{D}_{k+1}} = \ol{G}_{k}~T_{k}~\ol{G}_{k}^{\dagger} \ths\QEC\ths \ol{C}_{k+1} \ths T_{k+1} . \label{eq:easy_dressing}
	\end{gather}
	Using eq. \ref{eq:P_map_easy}, it is easy to see that $\ol{C^{D}_{k+1}}$ is equivalent to quantum error correction followed by an easy gate.

Figure \ref{fig:compiled_cycle} shows the result of compiling all of the twirling gates into the easy gates and quantum error correction routines. Note that the compiled circuit is logically equivalent to the original circuit in the absence of noise. However, in the presence of noise, the average output of the circuit is dictated by the performance of $\QEC(T)$ averaged over the different choices of Pauli gates $T$. This is what we refer to as $\QEC$ in the RC setting. In practice, this average performance can be achieved by repeating every iteration (shot) of the algorithm with a different Pauli operation compiled into the constituent $\QEC$ routines. In this paper, we have used the performance of the $\QEC$ routine under the twirled noise process as a proxy to the performance of $\QEC$ in the RC setting, for the analytical and numerical studies.

\def\vsepRCQEC{\vspace{0.3cm}}
\begin{figure}[h]
	\begin{center}
		\begin{subfigure}{0.9\textwidth}
			\centering
			\includegraphics[scale=\diagscale]{\diagramdir/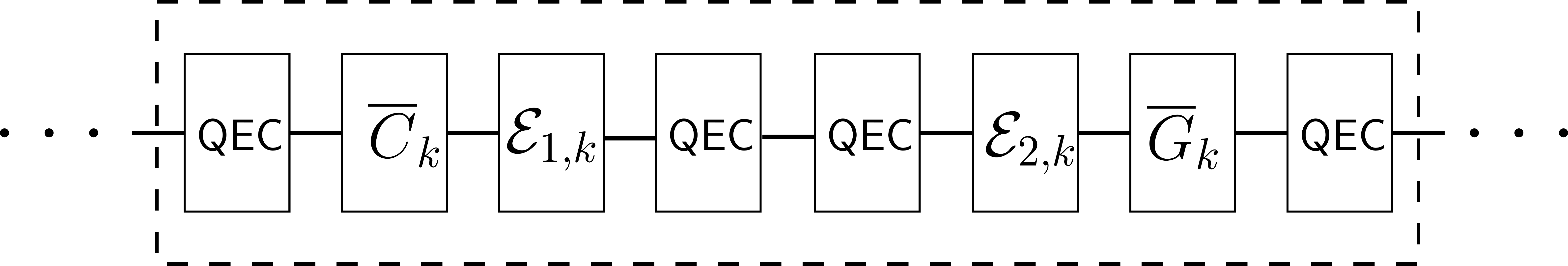}
			\caption{}
			\label{fig:clock_cycle}
		\end{subfigure}
		
		\vsepRCQEC
		
		\begin{subfigure}{0.9\textwidth}
			\centering
			\includegraphics[scale=\diagscale]{\diagramdir/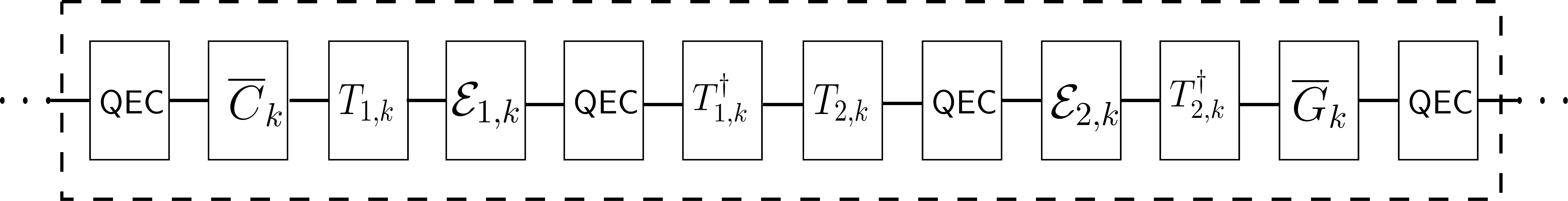}
			\caption{}
			\label{fig:insert_twirling}
		\end{subfigure}
		
		\vsepRCQEC
		
		\begin{subfigure}{0.9\textwidth}
			\centering
			\includegraphics[scale=\diagscale]{\diagramdir/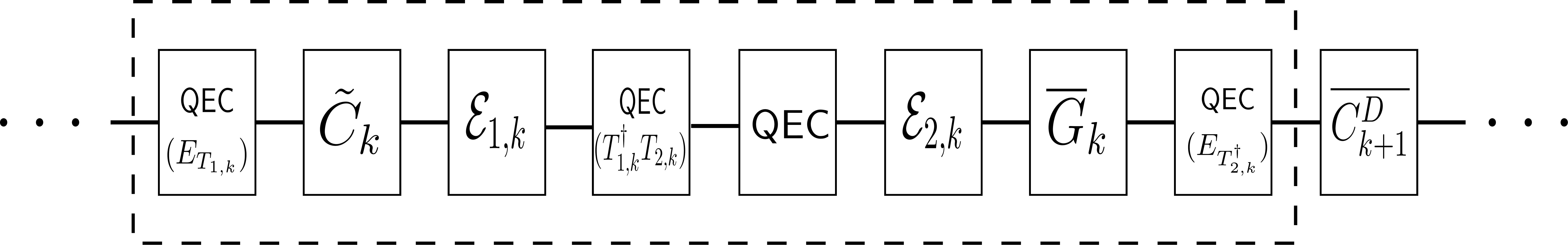}
			\caption{}
			\label{fig:compiled_cycle}
		\end{subfigure}
		\caption[RC in FT algorithm]{Figure \ref{fig:clock_cycle} shows the noisy gates in the $k-$th clock cycle of a fault tolerant quantum algorithm presented in the standard form prescribed in Ref.~\cite{WE16}. Twirling gates are inserted in fig. \ref{fig:insert_twirling} to tailor the noise processes to Pauli errors. These gates are compiled into existing gates by replacing easy gates by their dressed versions, in fig. \ref{fig:compiled_cycle}. The compiled circuit contains modified $\QEC$ routines with the twirling operators compiled into them.}
		\label{fig:rc_ftqec}
	\end{center}
\end{figure}

\end{widetext}
\end{appendix}
\end{document}